\documentclass[letterpaper,twocolumn,10pt]{article}
\usepackage{float}
\usepackage{usenix}

\usepackage{amsmath,amssymb}
\usepackage{xurl}
\usepackage{amsfonts}
\usepackage{array}
\usepackage{graphicx}
\usepackage{listings}
  \let\li\lstinline
\usepackage[frozencache]{minted}
  \setminted{fontsize=\footnotesize}
  \setmintedinline{fontsize=auto}

\usepackage{tikz}
  \usetikzlibrary{automata,positioning,arrows.meta,decorations.pathreplacing}
\usepackage{pgfplots}
  \pgfplotsset{compat=1.18}
  \usepgfplotslibrary{groupplots}
  \definecolor{clOSSL}{HTML}{4477AA}
  \definecolor{clRust}{HTML}{CC3311}
  \definecolor{clHACL}{HTML}{228833}
  \definecolor{clAWSLC}{HTML}{66CCEE}
  \definecolor{clFormosa}{HTML}{AA3377}
  \definecolor{clLweke}{HTML}{999999}
  \definecolor{clCrux}{HTML}{DDAA33}
  \definecolor{clOpus48}{HTML}{B07D2B}
  \definecolor{clSol}{HTML}{168AAD}
  \definecolor{clOpus5}{HTML}{C73E1D}
  \pgfplotsset{
    perfbar/.style={
      ybar, bar width=4.5pt,
      font=\scriptsize,
      enlarge x limits=0.35,
      ymin=0,
      ymajorgrids=true,
      major grid style={black!12, line width=0.3pt},
      ylabel style={font=\scriptsize},
      title style={font=\footnotesize\bfseries, yshift=-2pt},
      tick label style={font=\scriptsize},
      extra y ticks={100},
      extra y tick labels={},
      extra y tick style={grid=major, major grid style={dashed,black!40,line width=0.4pt}},
      every axis plot/.append style={fill opacity=0.85},
      legend style={font=\tiny, draw=none, fill=none, legend columns=-1,
        column sep=3pt, at={(0.5,1.02)}, anchor=south},
    },
  }

\usepackage{newunicodechar}
  \newunicodechar{⦄}{\ensuremath{|\!\}}}
  \newunicodechar{⦃}{\ensuremath{\{\!|}}
  \newunicodechar{∧}{\ensuremath{\wedge}}
  \newunicodechar{∀}{\ensuremath{\forall}}
  \newunicodechar{∃}{\ensuremath{\exists}}
  \newunicodechar{≠}{\ensuremath{\neq}}
  \newunicodechar{∨}{\ensuremath{\vee}}
  \newunicodechar{←}{\ensuremath{\leftarrow}}
  \newunicodechar{→}{\ensuremath{\rightarrow}}
  \newunicodechar{⟨}{\ensuremath{\langle}}
  \newunicodechar{⟩}{\ensuremath{\rangle}}
  \newunicodechar{⊢}{\ensuremath{\vdash}}

\usepackage{xspace}
\usepackage{setspace}

\usepackage{algorithm,algpseudocode}

\algrenewcommand\algorithmicdo{}

\usepackage[normalem]{ulem}

\begin{document}
\title{Scaling Verification of Cryptographic Software with Aeneas, Rust, and Lean\\[0.8ex]
  \large\mdseries Draft Technical Report, September 2026.}
\author{
{\rm Son Ho$^1$, Cédric Fournet$^1$, Jonathan Protzenko$^2$, Michael Naehrig$^{1}$}\\
{\rm Joshua Clune$^{1,3}$, Patrick Longa$^1$, Guillaume Boisseau$^4$, Fernando Leal Sánchez$^{4,5}$}\\
{\rm Aymeric Fromherz$^4$, Antoine Delignat-Lavaud$^1$}\\[0.5ex]
{\normalsize\rm $^1$Microsoft\quad $^2$Google\quad $^3$CMU\quad $^4$Inria\quad $^5$ENS Paris-Saclay}
}
\date{}

\newif\ifdraft

\ifdraft
  \DeclareRobustCommand{\son}[1]{ {\begingroup\color{red!60!black}{(Son) #1}\endgroup} }
  \DeclareRobustCommand{\jonathan}[1]{ {\begingroup\color{teal}{(Jonathan) #1}\endgroup} }
  \DeclareRobustCommand{\aymeric}[1]{ {\begingroup\color{blue}{(aymeric) #1}\endgroup} }
  \DeclareRobustCommand{\cedric}[1]{ {\begingroup\color{green!60!black}{(Cédric) #1}\endgroup} }
  \DeclareRobustCommand{\josh}[1]{ {\begingroup\color{orange!85!black}{(Josh) #1}\endgroup} }
  \DeclareRobustCommand{\michael}[1]{ {\begingroup\color{purple!85!black}{(Michael) #1}\endgroup} }
  \DeclareRobustCommand{\copilot}[1]{ {\begingroup\color{blue}{\emph{#1}}\endgroup} }
\else
  \DeclareRobustCommand{\son}[1]{}
  \DeclareRobustCommand{\jonathan}[1]{}
  \DeclareRobustCommand{\aymeric}[1]{}
  \DeclareRobustCommand{\cedric}[1]{}
  \DeclareRobustCommand{\josh}[1]{}
  \DeclareRobustCommand{\michael}[1]{}
  \DeclareRobustCommand{\copilot}[1]{}
\fi

\newcommand{\hacl}{\textsf{HACL$^\star$}\xspace}
\newcommand\fstar{F$^\star$\xspace}
\newcommand\lowstar{Low$^\star$\xspace}
\newmintinline[rustinline]{rust}{}
\newmintinline[leaninline]{lean}{}
\newmintinline[cinline]{c}{}

\newcommand{\totalintrinsics}{133}
\newcommand{\modelledintrinsics}{122}
\newcommand{\axiomintrinsics}{11}

\newcommand{\nat}{\mathbb{N}\xspace}

\newcommand{\myparagraph}[1]{\smallskip\noindent\textbf{#1}}

\newcommand{\sref}[1]{\S\ref{sec:#1}}
\newcommand{\slabel}[1]{\label{sec:#1}}
\newcommand{\fref}[1]{Figure~\ref{fig:#1}}
\newcommand{\flabel}[1]{\label{fig:#1}}


%


\maketitle

\begin{abstract}
We develop a new methodology for verifying
cryptographic software. 
We target production code written in Rust for performance and system integration, rather than verification convenience.
Rust's ownership discipline enables Aeneas to extract a pure model of this code in Lean, relieving us from low-level reasoning about pointer liveness and aliasing. Lean's extensibility lets us develop tactics and libraries that greatly simplify reasoning about extracted Rust code. 

We design and tune our toolchain to facilitate the use of AI.
Agents autonomously write formal proofs, which are independently verified by the Lean kernel.
Agents also assist in the formalization of cryptographic standards and platform-specific intrinsics, 
which still requires expert design and review. 

We apply our methodology to SymCrypt, Microsoft's cryptographic provider. We verify
its implementations of algorithms such as SHA-3 and ML-KEM, which were ported from C to Rust.
We also extend SymCrypt with experimental optimizations and implementations of algorithms such as FrodoKEM, ML-DSA, and HPKE to explore the scalability of writing, adapting, and verifying cryptographic code. 
Our 237~KLOC Lean development establishes safety, panic-freedom, and functional correctness of 16.7~KLOC of Rust code supporting post-quantum cipher suites
for x86-64 and ARM platforms.
Our evaluation shows that verified Rust 
can meet SymCrypt's performance, portability, deployment, and maintainability 
requirements. 
\end{abstract}


\section{Introduction}

Since HACL*~\cite{zinzindohoue2017hacl}, numerous projects have claimed
to provide verified cryptographic libraries that could be realistically
integrated into existing products and codebases. 
Ten years later, these claims mostly hold, but with two main caveats.

First, integration remains difficult.  
Whether it is Low*, Fiat-Crypto,
Vale, or Bedrock2, all of the toolchains used to verify low-level,
high-performance implementations have so far chosen to \emph{produce} code,
enabling tight control of the targeted subset and the shape of the verified code.
This makes sense from a scientific perspective: it is simpler 
to reason about a minimal ASM-like language 
equipped with decent control-flow (like Vale) rather than concrete assembly
syntax; or to reason about C code that relies only on \li+uintptr_t+ and pointers
to bytes (like Bedrock2) rather than the ill-defined C semantics of pointer
provenance, layout, and padding. 
Yet, engineers who have to do the actual 
integration work meet such generated code with defiance, 
and worry it may be nearly impossible to own, evolve and maintain for product teams.

Second, verification still entails significant friction:
it requires expertise, owing to the exotic nature of verification toolchains and the intense mental labor. 
Moreover, integrating and maintaining verified code in production remains challenging as code evolves continuously and forces frequent and tedious updates to the proofs.
Whether it is for assembly or C-like languages, the considerable overhead of memory reasoning, along with unpredictable proof automation, means verifying code is orders-of-magnitude harder than writing it.

In recent years, new tools have emerged that challenge these established tradeoffs.  
Our approach combines 
the safety and ownership guarantees of Rust (avoiding costly reasoning about memory),
the flexibility and metaprogramming of Lean (automating proofs), and 
the reasoning capabilities of AI agents (lowering initial and maintenance costs).
Together, these advances enable verification of code \emph{as written by the programmer} to keep up with code development.

\begin{figure}[t]
\centering
\includegraphics[width=1.0\columnwidth]{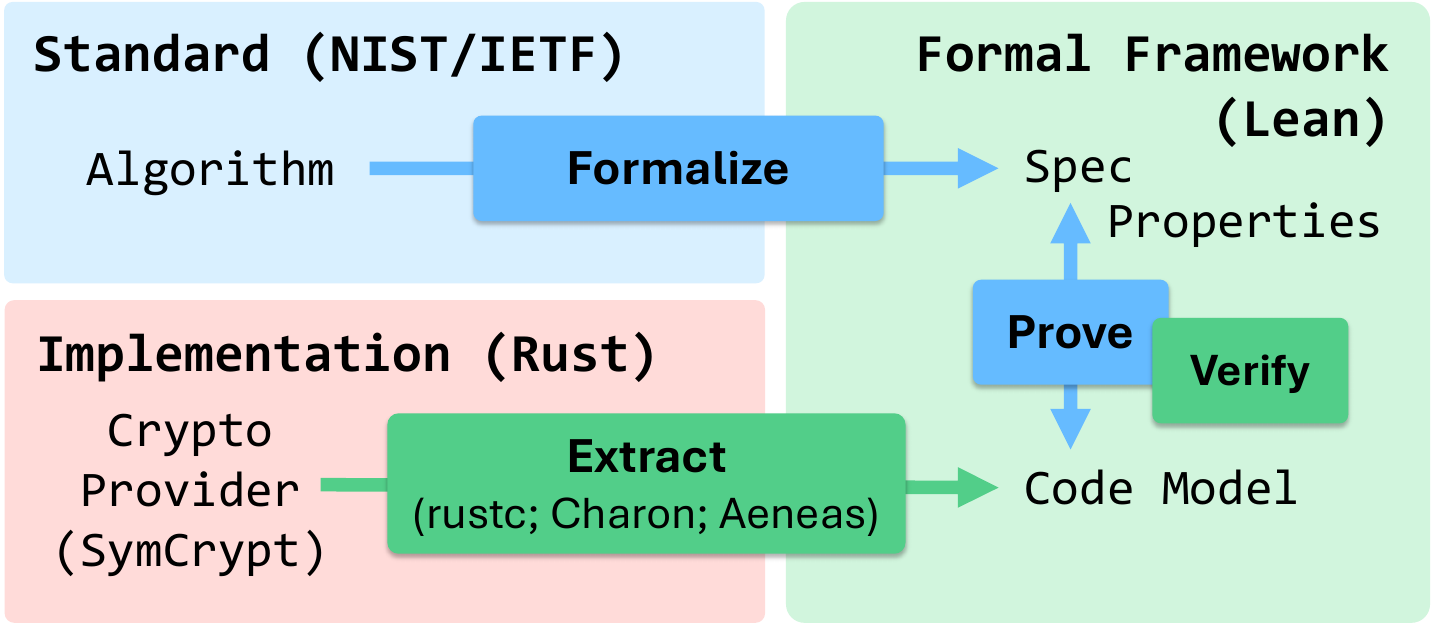}
\vspace{-2ex}
\caption{Overview of our verification methodology.
Green arrows: automated tools; blue arrows: agentic (best effort).}
\vspace{-1ex}
\label{fig:architecture}
\end{figure}

\myparagraph{Methodology and Toolchain (Figure~\ref{fig:architecture}).} 
At the specification level, we port standard NIST or IETF prose to the Lean theorem prover,
either manually or with the assistance of AI agents; we review this formalization
for alignment with the standard, test it, and equip it with high-level properties
to increase its trustworthiness.
At the implementation level, we use Charon~\cite{ho2025charon}, a driver for the Rust compiler, to extract rustc's Mid-level Intermediate
Representation (MIR) into a format called LLBC; we then use Aeneas~\cite{aeneas22} to translate LLBC into an equivalent Lean model,
leveraging the Rust type discipline to safely eliminate the details of low-level memory management.
Within the Lean framework, we finally prove that this model
correctly implements its specification.

To make this approach work on cryptographic code, 
we significantly extend Charon and Aeneas.
We add multi-target extraction to Charon, enabling the modular verification
of code that gets specialized for specific architectures at compile-time.
We extend the subset of Rust targeted by Aeneas by generalizing the handling of loops
and adding support for a large class of nested (mutable) borrows; this unlocks the
verification of common uses of iterators and iterator transformers.
Finally, we add a Lean backend to Aeneas with the accompanying libraries and
automation, notably for reasoning about arrays, slices, arithmetic, and bitvectors. 

\myparagraph{AI-Assisted Proof Engineering.}
Although we initially developed our methodology without AI assistance, 
agents became increasingly useful, and in turn 
motivated improvements in the toolchain. 
We delegate to autonomous agents the proofs about the extracted Lean model---and more generally the
generation of intermediate scaffolding such as lemmas, pre- and post-conditions, and loop
invariants---without adding AI machinery to the verification TCB since all proofs are independently verified by Lean. 

AI automation helps us efficiently verify Rust code as it is, keep formal specs closely aligned to the standards, 
and optimize theorems for clarity and ease of review, 
even when this incurs additional proof work. 
It relieves much of the tedium of maintaining a large formal development 
in sync with an active codebase, despite 
frequent code changes 
and tool updates---which typically require plenty of proof repairs. 

More generally, it shifts human focus to 
(1) increasing the quality of the relatively small trusted parts of the Lean development, 
namely the specification of standards and intrinsics and the main theorem statements 
attached to the verified code API; 
(2) extending and improving the toolchain to facilitate proofs, e.g., on arithmetic and bitvectors,
and handle bottlenecks, e.g., overly slow proofs for large Rust functions; and
(3) improving aspects of our implementations not related to verification, e.g.,
performance.

We report on our experience but do not attempt to quantify
productivity benefits, since the toolchain and agents significantly evolved throughout our project. 
For instance, it initially took us about six months to verify the NTT layer of ML-KEM whereas,
at the time of writing, it takes about
a week for an agent and a few days of human review
to formalize, implement, and verify an additional algorithm. 

\myparagraph{Case Study:} {SymCrypt~\cite{symcrypt}} is Microsoft's open-source cryptographic provider, supporting standard symmetric and asymmetric algorithms for a variety of x86 and ARM platforms. 
It is the primary cryptographic provider for Windows and Azure Linux (e.g., as an OpenSSL crypto provider).
While most algorithms are still coded in C and in assembly, it is gradually migrating to Rust, notably for new post-quantum algorithms. 

We verified its deployed Rust implementations, such as SHA-3 and ML-KEM,
finding minor flaws in the process.
We also developed experimental optimizations 
and new implementations for additional algorithms, such as ML-DSA, FrodoKEM, and HPKE, 
with two research goals: 
scale verification to include code that composes existing primitives and exports 
a broader API for performance evaluation; 
and explore joint agentic coding and proving for new optimizations and new algorithms.
(Although verified, integrated, and tested, we do not claim our experimental code is production-ready.)
In combination, this Rust code supports, for instance, post-quantum-secure TLS 1.3 ciphersuites, and is generally faster than its C counterpart. 

\myparagraph{Main Contributions} 
\begin{enumerate}
\item A methodology for verifying production code written in Rust
using a companion Lean formalization.

\item Extending Aeneas to scale verification
  and efficiently reason about Rust code 
  in a new Lean backend.

\item Formalizations of cryptographic standards 
(featuring specification-level properties and test vectors) and of intrinsics for platform-specific code.

\item Verifying the safety, correctness, and panic-freedom of SymCrypt Rust code, notably for post-quantum cryptography, as well as fresh experimental code. 

\item Performance evaluation versus C/C++ crypto providers and other verified implementations; evaluation of AI-assisted proof costs and integration with SymCrypt.

\end{enumerate}

\myparagraph{Contents.} 
The paper is structured as follows.
\sref{background} gives background on Lean and Aeneas.
\sref{trust} states our trust assumptions and verification scope.
\sref{specs} explains how cryptography is formalized. 
\sref{methodology} describes our extensions to Aeneas, its Lean backend, and the accompanying tactics and libraries.
\sref{intrinsics} explains how we model and validate platform-specific intrinsics.
\sref{casestudy} presents the SymCrypt case study and illustrates selected verification patterns.
\sref{evaluation} evaluates the use of agents, integration, and performance of the resulting code.
\sref{related} discusses related work and concludes.

\section{Background}
\slabel{background}

\myparagraph{Lean} is a theorem prover 
combined with a dependently-typed language, used to formalize 
specifications and prove their properties.
Tactics manipulate goals 
and produce proof terms checked by Lean's small kernel.
Lean displays current hypotheses and goals, enabling 
interactive proof development. 
It provides powerful automation, e.g., an SMT-like \leaninline{grind} tactic, 
and metaprogramming to extend its syntax, commands, and tactics.
Lean has been widely applied to mathematics~\cite{mathlib2020} and
program verification~\cite{mvcgen,loom,cedar2024,diffPrivacy2024,iris-lean}.
It is the focus of much AI effort, with agents 
demonstrating strong abilities at mechanizing proofs autonomously.

\myparagraph{Aeneas~\cite{aeneas22}} is a verification framework for Rust.
Like other toolchains~\cite{creusot22,fluxPLDI22,verusSOSP24,thrustPLDI25,ullrich2016electrolysis,bhargavan2024hax},
it leverages Rust's system of ownership and borrows to encode programs to 
pure models with no notion of pointers or memory.
Charon extracts rustc's MIR into LLBC, which Aeneas consumes
by translating every borrow, whether mutable or shared,
to a value without relying on state-passing or mutability.
Aeneas emits these functional models for different verification frameworks,
allowing users to prove theorems that apply to the original Rust programs.
This \emph{extrinsic} style of verification clearly separates
the implementation (in Rust) and its formal development, 
instead of inserting specifications and proofs directly in the source code.

\section{Trust Assumptions}
\slabel{trust}

Our formal theorems establish functional correctness and panic-freedom for the Rust
code, against the formalized standards, and can be independently re-checked by
the Lean kernel. This leaves three questions for review: is it the right formal
specification, is it the right set of properties for this code, and what trust
assumptions do they rely on? We summarize below the main assumptions our results
entail.

In general, we trust the Rust toolchain, notably its compiler and its
platform-specific back-ends for x86-64 (including SSE2/AVX2) and AArch64
(including NEON). We also trust that code using our verified implementations is safe,
complies with its pre-conditions, and is compiled for the targets
covered by our verification (x86-64 and AArch64).
We trust the runtime platform meets the
toolchain's expectations; in particular, although the Rust code is defensively
written in constant-time style to protect against microarchitectural side
channels, we verify only functional correctness, not leakage resistance. We
trust that Charon correctly extracts MIR into LLBC and that Aeneas correctly translates LLBC into Lean,
the soundness of the Lean kernel that checks all proofs in our
development, and the soundness of the compiled \li+native_decide+ and \li+bv_decide+ tactics.
Finally, we trust the soundness of the Aeneas axiomatization for some functions of the
Rust standard library, and in particular the axiomatization of intrinsics, which we
carefully transcribe from hardware-vendor reference documents.

For each algorithm, we trust the Lean formal specification faithfully captures the standard, 
hence requiring careful human review and testing.
We also trust that the verified pre- and post-conditions of the public interface
fully capture the intended functional behavior, which can likewise be reviewed by
inspecting the Lean theorems.
Conversely, we need not trust the Lean proofs, nor the lower-level helpers,
properties, and intermediate specifications attached to the code (intermediate
pre- and post-conditions, loop invariants, range hypotheses, etc.), since they
are all systematically checked by Lean. This allows us to delegate most of the
proof work to AI agents without having to trust their output.

Side channels are out of scope at the Rust level: `constant-time' code 
is verified for correctness but not leakage protection.
Instead, SymCrypt uses leakage-specific fuzzing 
and symbolic execution on compiled binaries, which provides 
lower-level informal protection against known microarchitectural side channels 
on x86-64 platforms. 
 
\section{Formalizing Crypto Standards}
\slabel{specs}

Specifications are the cornerstone of any verification effort, and their quality is
critical: a formal proof provides limited value if the final theorem
does not correctly capture what the program is intended to do.
Starting from high-quality standards, we develop 
formal specifications optimized for clarity, fidelity, brevity, and ease of review.

\myparagraph{Mechanizing Standards.} To verify the functional correctness of
cryptographic implementations, we consider the standard the source of truth and
aim to minimize the gap between its text and its transcription in Lean.
We develop specifications independently of the Rust implementation and its
correctness proofs; this avoids introducing implementation-specific details
merely to simplify proofs.
For example, if a standard uses bitstrings, so should the specification (not byte arrays).

We strive to syntactically align specifications and standards,
ideally line-by-line, symbol by symbol, so that we can review them side by side;
for instance, we show in \fref{ntt-spec} our formalization of ML-KEM's NTT algorithm in Lean.
As standards use pseudocode that leverages imperative features such as in-place updates
and loops, we leverage Lean's monadic do notation to port them as is.
We do not compromise on the specification by adapting it when it could be simplified.
Similarly, when a loop in the standard could be simplified to, e.g., a call to a
\leaninline{map} function, we preserve the loop in the specification and prove an
auxiliary theorem on the side.

\myparagraph{Linking specifications.}
Whenever a standard functionally depends on another, we link their formal specifications accordingly.
For example, the ML-KEM specification calls the SHA-3 extensible hash specification, 
using a thin adapter to bridge their notations. 
Proving that the implementation of ML-KEM is correctly wired to the implementation of SHA-3 is thus in scope for
verification.

\myparagraph{Reusing Mathematical Libraries.}
Our specifications use Lean's vast libraries of mathematical definitions and results, 
for example libraries for finite fields, linear algebra, and polynomial extensions. 
This reduces custom formalization and provides well-developed definitions,
theories, and lemmas, facilitating both review and proof development.

\begin{figure}[t]
\setcounter{algorithm}{8}
\setlength{\intextsep}{0pt}
\begin{algorithm}[H]
\small
\caption{$\textsf{NTT}(f)$}
\label{alg:ntt}
\begin{algorithmic}[1]
\Require array $f \in \mathbb{Z}_q^{256}$
\Ensure array $\hat{f} \in \mathbb{Z}_q^{256}$
\State $\hat{f} \gets f$
\State $i \gets 1$
\For{($\mathit{len} \gets 128$; $\mathit{len} \geq 2$; $\mathit{len} \gets \mathit{len}/2$)}
  \For{($\mathit{start} \gets 0$; $\mathit{start} < 256$; $\mathit{start} \gets \mathit{start} + 2 \cdot \mathit{len}$)}
    \State $\mathit{zeta} \gets \zeta^{\text{BitRev}_7(i)} \bmod q$
    \State $i \gets i + 1$
    \For{($j \gets \mathit{start}$; $j < \mathit{start} + \mathit{len}$; $j\raisebox{0.2ex}{\scriptsize \(++\)}$)}
      \State $t \gets \mathit{zeta} \cdot \hat{f}[j + \mathit{len}]$
      \State $\hat{f}[j + \mathit{len}] \gets \hat{f}[j] - t$
      \State $\hat{f}[j] \gets \hat{f}[j] + t$
    \EndFor
  \EndFor
\EndFor
\State \Return $\hat{f}$
\end{algorithmic}
\end{algorithm}
\begingroup
\setlength{\topsep}{0pt}
\setlength{\partopsep}{0pt}
\renewcommand{\theFancyVerbLine}{%
  \ifnum\value{FancyVerbLine}=0\relax\else\normalfont\footnotesize\arabic{FancyVerbLine}:\fi}
\vspace{-0.5\baselineskip}
\begin{minted}[escapeinside=||,linenos,firstnumber=0,fontsize=\footnotesize,
  xleftmargin=1.7em,numbersep=0.5em]{lean}
def NTT (f : Polynomial) : Polynomial := Id.run do
  let mut «|$\hat{f}$|» := f
  let mut i := 1
  for h0: len in [128 : >1 : /= 2] do
    for h1: start in [0 : 256 : 2*len] do
      let zeta := |$\zeta$| ^ (bitRev 7 i)
      i := i + 1
      for h: j in [start : start+len] do
        let t := zeta * «|$\hat{f}$|»[j + len]
        «|$\hat{f}$|» := «|$\hat{f}$|».set (j + len) («|$\hat{f}$|»[j] - t)
        «|$\hat{f}$|» := «|$\hat{f}$|».set j         («|$\hat{f}$|»[j] + t)
  return «|$\hat{f}$|»
\end{minted}
\vspace{-1.1\baselineskip}
\endgroup
\noindent\rule{\linewidth}{0.4pt}
\vspace{-4ex}
\caption{ML-KEM's NTT specification in the FIPS-203 standard (top)
and its Lean formalization (bottom)}
\flabel{ntt-spec}
\end{figure}

\myparagraph{Hardening with Tests.}
Our Lean specifications are executable, albeit much slower than
their Rust counterparts. 
We test them against official known-answer test (KAT) vectors, 
to detect misunderstandings or transcription errors during formalization.
Later on, differential testing with, e.g., Rust implementations
also helps investigate any discrepancy.
This effort is independent of larger-scale implementation testing, 
which is still very much required to gain confidence 
for properties outside the scope of our formal model.

\myparagraph{Hardening with High-Level Properties.}
Specifications are additionally equipped with high-level mathematical properties (and
proofs). 
We aim to cover the properties mentioned in the standard
(e.g., Keccak is a permutation), the standard's internal consistency
(e.g., tables of constants are correct), 
functional correctness properties involving related algorithms 
(e.g., decryption is the inverse of encryption), and the correctness of the specification 
itself against its mathematical foundation.

As an example, take the number-theoretic transform (NTT) used in ML-KEM (and similarly in
ML-DSA). The FIPS standard~\cite[\S4.3]{fips203} informally describes the
NTT as a ring isomorphism for fast polynomial multiplication.
However, verifying that the standard's Algorithm~9, which gives the pseudocode for the NTT,
actually computes this isomorphism requires a precise understanding of the
algebraic structures and their algorithmic realization.
Another example concerns the decryption failure probabilities of
ML-KEM and FrodoKEM~\cite{frodokem}, whose
specifications provide upper bounds stating that the chosen
parameter sets keep failure rates below negligible values such that they do not invalidate
the scheme's security claims.
Verifying the claimed values in the specification requires an understanding of the
underlying error distributions, how errors propagate and accumulate through various
operations, and carrying out a precise estimation.
Our formal development proves that the specified NTT realizes the intended ring
isomorphism and establishes the claimed decryption-failure bounds,
leveraging Mathlib's associated concepts and theorems.

\section{Rust Verification with Aeneas}
\slabel{methodology}

\subsection{Extraction to Lean}\label{sec:reduce}

Our verification methodology crucially relies on Aeneas \cite{aeneas22,ho2024} to
translate the MIR of a large class of stateful Rust programs into pure, side-effect-free functions that
represent mutations by explicitly threading updated values.
We illustrate the translation on a function (simplified from ML-KEM) that 
applies a Barrett reduction in-place via a mutable borrow:

\begin{minted}{rust}
const Q: u64 = 3329;      // prime used in ML-KEM
const R: u64 = 1_290_168; // ceil(2^32 / Q)

// returns x % Q when x < 2*Q
fn reduce(x: &mut u64) {
  let quotient = (*x * R) >> 32;
  *x = *x - quotient * Q; }
\end{minted}

\noindent 
which Aeneas translates into a corresponding Lean function
that explicitly returns the reduced value:

\begin{minted}{lean}
def reduce (x : U64) : Result U64 := do
  let y ← x * R
  let quotient ← y >>> 32#i32
  let z ← quotient * Q
  x - z
\end{minted}

Rust code may panic for a variety of reasons: in our example, 
the program fails if any of these steps causes a \leaninline{U64} overflow or underflow.  
Aeneas makes this explicit, by returning a Lean \leaninline{Result} and using Lean's monadic do-notation: 
each monadic let binding (denoted by ←) 
either explicitly propagates failure or unwraps a value
and continues with the next step.  
Aeneas also introduces intermediate steps    
so that every fallible operation has its own monadic binding.

In more complex cases, such as a Rust function 
that returns a mutable borrow, 
Aeneas would generate a \emph{backward function}, which the caller invokes 
when the borrow ends to propagate the final borrowed value back to the original owner. 

More generally,
Aeneas supports common uses of nested borrows, loops (translated
to recursive functions), traits (translated to structures), and mutable
iterators; these suffice for our cryptographic implementations.
On the other hand, Aeneas has limited support for unsafe Rust and does not support raw pointers, interior mutability, or concurrency;
functions using unsupported features must be modelled by hand. 

Finally, Aeneas applies a sequence of transformations designed to keep the generated Lean
close to the original Rust.
The resulting definitions form the extracted model that we verify with the Lean proof framework described next.

\subsection{Reasoning about Monadic Code}
\slabel{monadic}

We also extend Aeneas with Lean commands, tactics, and libraries for reasoning about the
extracted Lean model; the resulting backend consists in 29k lines of Lean.
Continuing with our example, they can be used to state and prove that \leaninline{reduce}
is functionally correct and panic-free, as follows: 

\slabel{reduce.spec}
\begin{minted}[linenos,numbers=left,numbersep=-5pt,fontsize=\footnotesize]{lean}
  @[step]
  theorem reduce.spec 
    (x : U64) 
    (h : x.val < 2 * Q.val) :
      reduce x ⦃ (r : U64) => r.val = x.val % Q.val ⦄ := by
    unfold reduce 
    step* <;> scalar_tac
\end{minted}

In this theorem, 
\leaninline{x} is the initial value of the borrow, 
\leaninline{h} is a precondition on \leaninline{x}, 
and \leaninline{⦃ r => ... ⦄} is a monadic postcondition on
the final value of the borrow, named \leaninline{r}. 
Hence, the theorem states that, 
if \leaninline$x$, interpreted as an (unbounded) natural number \leaninline$x.val$, 
is smaller than \leaninline{2*Q}, 
then \leaninline{reduce x} succeeds (no panic) and returns 
the correct value \leaninline{r = x 
interpreted as a natural number. 
The theorem is marked with a \leaninline{@[step]} attribute, 
to save it in a database of code properties, so that
proofs of functions that call \leaninline{reduce} can use 
it as a lemma without having to unfold its definition. 

The proof (lines 6--7) first unfolds the body of the \leaninline{reduce} function, 
listed above, then applies the \leaninline{step*} tactic  
to step through its monadic let bindings. 
In our case, \leaninline{step*} leaves three arithmetic goals to be proved:
(1) the second multiplication does not overflow; 
(2) the subtraction does not underflow; and 
(3) the result satisfies the postcondition, 
and all these goals are automatically discharged by the subsequent
\leaninline{scalar_tac} tactic.
We show below the last intermediate proof goal, as displayed by the interactive proof mode:
\begin{minted}[mathescape,fontsize=\footnotesize]{lean}
... -- omitting a few hypotheses
h : x.val < 2 * Q.val
y_post : y.val = x.val * R.val
quotient_post : quotient.val = y.val >>> 32
z_post : z.val = quotient.val * Q.val
r_post : r.val = x.val - z.val
⊢ r.val = x.val % Q.val
\end{minted}
Thanks to prior simplifications, this goal captures 
the correctness of our function, implementing the modulo
reduction with two multiplications, a shift, and a subtraction,
purely in terms of integer arithmetic,
which is easily handled by Lean.

We now explain in more detail our new Aeneas tactics for proving monadic postconditions.
The \leaninline{step} tactic moves one monadic step forward, 
attempting to prove preconditions on the fly by means of \leaninline{grind} 
(Lean's SMT-like tactic)
with the goal of providing an experience akin to step debugging. 
For example, \leaninline{step} handles the first monadic step above (\leaninline{y ← x * R})
by applying the Aeneas specification of Rust \leaninline{U64} multiplication, 
automatically proving its precondition (no overflow, using hypothesis \leaninline{h}), 
and adding to the proof 
context the resulting integer equation \leaninline{y_post : y.val = x.val * R.val}. 
The \leaninline{step*} tactic repeatedly applies \leaninline{step} while
performing case disjunctions on the fly. 
Finally, the \leaninline{step*?} tactic
expands into a detailed proof script,
allowing the user to inspect intermediate lemmas and remaining proof goals.

These tactics are similar to those provided by other verification frameworks in Lean~\cite{mvcgen,loom} but carefully tuned to extracted Rust code. 
To improve usability, the \leaninline{step} tactic provides features to control names
(with syntax \leaninline{step as ⟨ ... ⟩}), infer ghost variables,
and provide custom tactics to discharge preconditions when the default \leaninline{grind} does not suffice.
Following standard practice from verifiers that leverage SMT-like
automation~\cite{verusSOSP24,Ironfleet}, we fine-tune calls to
\leaninline{grind} to strike a trade-off between speed and
automation; in particular, to leverage Lean's interactivity,
we aim for \leaninline{grind} to fail fast when a goal cannot be closed,
rather than attempt a wide range of heuristics.
To this end, we use a carefully curated set of lemmas for e-matching
and keep all \leaninline{grind} parameters low by default (e.g., number of case splits, maximum
number of generations allowed when instantiating theorems, etc.).
When using \leaninline{step*}, we also carefully thread the state accumulated by
\leaninline{grind} to avoid repeatedly deriving the same facts, thus keeping the proof
time low. 

\subsection{Reshaping Code for Ease of Verification}

In classic \emph{proof-oriented programming}, code and specs can be 
written to facilitate verification, e.g.,
by avoiding patterns not supported by the tools or awkward to verify, 
or by splitting large code into auxiliary functions, each aligned 
with a logical part of the specification.

With our methodology, verification starts from existing code and independently formalized standards. 
Instead of adapting Rust source code, we leverage Lean support for 
extrinsic verification to reshape code \emph{after extraction}.
As a basic example, we rely on Lean's simplifier and generic equality lemmas 
to selectively unfold loops. 
Leveraging Lean's extensibility and metaprogramming facilities, 
we also develop a new Lean command, \leaninline{decompose}, to automatically fold 
selected code parts into auxiliary function calls. 
Our command takes as input a target function 
and a pattern-matching description of the transformations to apply to its body. 
The command defines auxiliary functions for the folded code (or re-uses predefined functions) 
together with a theorem that proves the transformations preserve functional equality. 
Importantly, the command and its use are outside the verification TCB---the resulting 
proofs are checked by the Lean kernel, as usual. 

Consider, for example, a function with two inlined copies of the \leaninline{reduce} function in~\sref{reduce}, of the form (in Lean):

\begin{minted}[linenos,numbers=left,numbersep=-5pt]{lean}
def compute (a b : U64) : Result U64 := do
  let y0 ← a * R
  let q0 ← y0 >>> 32#i32
  let z0 ← q0 * Q
  let a0 ← a - z0
  ... 
\end{minted}
where we omit the second copy to reduce argument~\leaninline{b} and the rest of the code.
We can fold both copies with the command:
\begin{minted}{lean}
#decompose compute compute_eq 
  letRange 0 4 => reduce
  letRange 1 4 => reduce
\end{minted}
that replaces, e.g., lines 2--5 (positions 0--3 in the Lean AST) with \leaninline{let a0 ← reduce a}.
The command also defines (and proves) a theorem \leaninline{compute_eq} 
that equates the main function body before and after rewriting. 
This enables a simpler proof of \leaninline{compute}: 
we apply the tactics \leaninline{rw [compute_eq]; step*} 
to simplify the function body, then step through the 
rewritten code, automatically applying theorem \leaninline{reduce.spec} of \sref{reduce.spec} 
to establish that both arguments are correctly reduced
before dealing with the rest of the code. 

The \leaninline{decompose} command 
is particularly useful to handle large arithmetic functions (see~\sref{casestudy}), 
notably those produced by Rust macros (expanded before the MIR is generated). 
While not natively provided by Lean, it is effectively used by agents, which are instructed to apply it 
on large code (above a certain threshold of monadic steps) as well as code whose monolithic proof 
turns out to be too complex to automate. 
Their use also led to successive improvements to the command.

\subsection{Domain-Specific Automation}

Scaling verification to cryptographic implementations
requires powerful automation, in particular for monadic code 
that combines arrays, (non-linear) arithmetic, and bitwise operations.
Lean provides native automation, for instance with its SMT-like solver \leaninline{grind},
and its bit-blaster \leaninline{bv_decide},
but they still sometimes fall short.
We therefore implement our own tactics,
tuned to the code we extract from Rust.

The new tactics \leaninline{simp_scalar} and 
\leaninline{simp_lists} perform conditional
simplifications such as \leaninline{a 
\leaninline{(l.set i x)[j] = l[j]} if \leaninline{i ≠ j}.
They are roughly equivalent to calling \leaninline{simp} with \leaninline{scalar_tac} as a discharger,
but carefully optimized for Lean elaboration performance, 
efficiently massaging goals so that \leaninline{grind} can conclude, 
and chaining sequences of otherwise manual simplifications.

Other new tactics similar to Mathlib's \leaninline{zify}
convert goals and hypotheses between machine integers, natural
numbers, integers modulo \leaninline{p} (\leaninline{ZMod p}), and bit vectors, 
enabling, e.g., \leaninline{bv_decide} to conclude. 
This requires discharging arithmetic obligations on the fly in a manner similar to
\leaninline{simp_scalar};
we use the resulting conversion tactic to implement \leaninline{bv_tac},
a tactic that lifts the context before calling \leaninline{bv_decide}, saving
many steps when closing bitvector goals.

\subsection{Modelling Rust Standard Libraries}
\slabel{stdlib}

Our Lean backend provides a collection of models for Rust primitives and standard libraries,
covering the needs of our target code.
For instance, we embed the Rust type \rustinline{U32} 
as the Lean subtype 
\leaninline[escapeinside=||]{{x : Nat // x |$\leq$| U32.max}},
that is, natural numbers less than or equal to \leaninline{U32.max},
and the Rust type \rustinline{[T; N]} of fixed-length arrays 
as a subtype of \leaninline{List}.
We translate \rustinline{assert!}
to a function \leaninline{massert} that fails
if its argument is false, thereby requiring that 
all dynamic assertions in Rust be statically discharged as part of verification.
We also provide models, e.g., for Rust iterators and iterator transformers. 

To increase confidence in these models, we write tests in Rust,
translate them with Aeneas, and (leveraging the fact that the generated Lean definitions are executable)
run them in both Rust and Lean to compare their results.
This involves a custom \rustinline{verify::test} attribute
that instructs Aeneas to generate a unit test for the extracted function. 

We axiomatize functions not currently supported by the translation,
using the \rustinline{verify::opaque} attribute 
to instruct Aeneas to emit only their type signature,
and writing instead a trusted model in Lean. 
We carefully review such models and test them whenever possible.
For example, SymCrypt calls a function \rustinline{wipe_slice} 
to clear private state, implemented via raw pointers and 
external code to resist compiler optimization.  
We mark it as opaque and provide an axiom 
that reflects its specification: setting all elements of the slice to 0.  


\subsection{Delegating Verification to Agents}

We write skill files to provide technical documentation and examples about Aeneas extraction, commands, tactics, and libraries, for example 
explaining when and how to use the \leaninline{decompose} command, how to handle excessively slow proofs,
and how to repair proofs following an Aeneas update. 

We structure AI automation around 
\emph{verification campaigns}, prescribing the sequence 
of stages agents must follow to verify a given implementation against 
its specification.
Each stage defines the files in scope, e.g., 
agents cannot edit specifications, axioms, and implementations during proof development: this 
requires escalation and explicit human approval. 
Each stage is gated by reviews by humans and independent agents, 
with detailed requirements to be met to mark them complete. 

Campaigns roughly go as follows: 
(1) Formalize the standard; see \sref{specs}.
(2) Survey the Rust code, its API, its tests, and its extraction to Lean.
(3) Scaffold the proof, writing a theorem statement for every Rust function,  
as well as pre- and post-conditions, and loop invariants; all proofs are admitted at this stage---the goal is to prevent backtracking while developing proofs, e.g., because a precondition is missing.
The end of this stage is gated on an API review (aka the "main theorems")
to ensure their statements are as simple and complete as possible. 
(4) Complete the proofs; this stage is the most costly but it can be parallelized. 
\sref{eval-ai} provides additional experimental information. 

\subsection{Multi-Target Verification}
\slabel{multitarget}

The semantics of a Rust program
depend on the target architecture for which it is compiled.
This can be implicit, via constants like \rustinline{usize::BITS} 
that can leak into the MIR because of rustc's constant evaluation,
or explicit, via \rustinline{cfg} attributes in source code 
to statically select between target-specific implementations. 
The latter case is particularly common in fast cryptographic implementations
(see~\sref{intrinsics} and \sref{casestudy}).

A challenge is that rustc inherently targets one architecture at a time: the MIR it outputs loses all information about earlier compile-time specialization. 
Left unchecked, this would require replicating our whole verification effort for each target.

Instead, we extend Charon so that it invokes rustc once for each target
in scope for joint verification, then merges the resulting LLBC ASTs.
The vast majority of the code is unchanged and deduplicated.
The few functions that are target-specific are merged, turning static branches into dynamic branches.
Appendix~\ref{sec:nttarch} illustrates multi-target extraction and verification 
for the different backends of the ML-KEM NTT.

\section{Modelling Target-Specific CPU Features}
\slabel{intrinsics}

Cryptographic libraries owe much of their performance to specialized CPU instructions,
comprising both general-purpose SIMD instructions to increase throughput, and custom 
instructions, such as AES-NI to accelerate AES.
Rust surfaces these instructions as intrinsics that we need to precisely model.
Intrinsics complicate code (and its verification) because they lack portability and are potentially unsafe.
Next, we recall what Rust provides, explain how SymCrypt uses intrinsics, and describe their modelling in Rust and Lean.

\myparagraph{Intrinsics in Rust.}
Rust exposes instructions through the \rustinline{core::arch} module,
which provides native functions such as \rustinline{_mm256_add_epi64} for
AVX2 and \rustinline{vaddq_u32} for NEON, operating on opaque vector types such as
\rustinline{__m256i} and \rustinline{uint8x16_t}. 

Three language mechanisms control the use of intrinsics in code.  
(1) Compile-time target architecture, e.g. \rustinline{#[cfg(target_arch = "x86_64")]}, 
includes or excludes whole modules, so x86-64 and AArch64 back-ends never coexist in a binary. 
(2) Static feature gating, e.g. \rustinline{#[target_feature(enable = "avx2")]},
asserts that a set of intrinsics is available to compile a function
and thus enable their safe use in its code, but---since running that function on a CPU
lacking the feature is undefined behavior---the function itself becomes
\rustinline{unsafe}, making it the caller's responsibility to ensure they are actually present.
A feature in the target's baseline ABI (e.g. SSE2 on \rustinline{x86_64}) is guaranteed
everywhere, so its intrinsics are callable from safe code with no annotation. 
(3) Dynamic feature detection, e.g., the standard-library macro \rustinline{is_x86_feature_detected!}, 
queries the CPU at runtime; successful detection justifies entering an \rustinline{unsafe} block that calls 
the corresponding \rustinline{target_feature} function. 
The standard library provides
the intrinsics' semantics only by reference to the vendor manuals, not as Rust-level
specifications.

\myparagraph{Intrinsics in SymCrypt.}
SymCrypt combines all three mechanisms. For the NTT kernel of ML-KEM, for example, 
it defines a trait (\rustinline{NttIntrinsics}) whose methods abstract the lane operations
the algorithm employs, with one instance per back-end
(\rustinline{NttIntrinsicsXmm}, \rustinline{NttIntrinsicsNeon}).
The kernel is written as a generic function
\rustinline{ntt_layer::<T: NttIntrinsics>}, which
the compiler monomorphizes for each back-end.
A small entry function then selects a back-end, 
with \rustinline{target_arch} arms to select which back-ends are even compiled,
conditional branches for back-ends requiring optional features (explained next), 
and portable code as fallback.

SymCrypt implements dynamic feature detection in C. 
At initialization, it sets flags for each feature, 
depending on hardware detection and load-time configuration.
Rust code tests them using the inlined function \rustinline{cpu_features_present},
which we model as follows: 
\begin{minted}{lean}
axiom featurePresent : U32 -> Bool
@[step] axiom cpu_features_present.spec (feat : U32) :
  cpu_features_present feat ⦃ r => r = featurePresent feat ⦄
\end{minted}

To verify the safety of using these features, we require, e.g., \leaninline{featurePresent FEATURE_BMI2}
(where \rustinline{FEATURE_BMI2} is SymCrypt's constant for the BMI2 extension)
as a precondition of every function that may use BMI2 instructions, 
either explicitly via intrinsics, or implicitly via \rustinline{#[target_feature(enable = "bmi2")]}.

Note that feature presence in Lean is both fixed and opaque (i.e., axiomatized); this matters
for verifying representation invariants that depend on a specific feature (e.g. PCLMULQDQ) 
and that must be preserved across multiple calls to code 
that branches on \rustinline{cpu_features_present} (e.g. for GCM).

\begin{figure}[t]
\centering\small
\begin{tabular}{| l l | r r |}
\hline
Arch & Extension & Rust body & Opaque shim \\
\hline
---      & portable lane ops &  20 & 0 \\
---      & shared AES        &   0 & 4 \\
x86-64   & SSE2              &  32 & 0 \\
x86-64   & SSSE3             &   2 & 0 \\
x86-64   & AVX2              &  23 & 0 \\
x86-64   & AES-NI            &  13 & 3 \\
x86-64   & SHA-NI            &   2 & 3 \\
x86-64   & PCLMULQDQ         &   0 & 1 \\
AArch64  & NEON              &  26 & 0 \\
AArch64  & Armv8 AES         &   4 & 0 \\
\hline
         & Total             & \modelledintrinsics{} & \axiomintrinsics{} \\
\hline
\end{tabular}
\caption{Verification shims for hardware intrinsics, 
either with a verified Rust body or marked as opaque 
and axiomatized in Lean. The AES row
factors out axioms used by both x86-64 
and AArch64 shims.}
\label{tab:intrinsics}
\end{figure}

\myparagraph{Modelling Intrinsics in Rust and Lean.}
Using the \rustinline{verify} feature, 
we statically redirect every intrinsic call to a \emph{verify-only shim} in a
separate \rustinline{src/verify/intrinsics/} subtree. These files are never
compiled in a production build. 

Each shim is a Rust function over fixed-size lane arrays, e.g. 
\rustinline{[u32; 4]} and \rustinline{[u64; 2]} for 128-bit registers. 
In most cases, the shim provides a Rust model: 
its body is a line-by-line transcription of the vendor 
\emph{Operation} pseudocode into ordinary lane operations, 
which Aeneas translates into a pure Lean function
we then verify against its lane specification (as any other function).
In the few remaining cases, we write instead a Lean axiom. 
For example, when the vendor documentation states
that an intrinsic implements an AES or SHA2 round, 
a Rust transcription would not bring much, so we mark the shim
\rustinline{#[verify::opaque]} and provide an axiom 
that refers directly to the formalization of the corresponding 
standard. 
The two cases yield the same form of Lean statements, enabling the functional verification 
of code that calls the intrinsics, and both involve a trust assumption. 
The main difference is that those given a Rust model are theorems instead of axioms,
and that they are easier to review and test. 
Hence, we can use our models as an alternative (slower) implementation 
to run existing Rust tests on SymCrypt algorithms.
Appendix~\ref{sec:intrinsics-examples} provides examples illustrating these two cases.

Each shim, modelled or axiomatized, is carefully reviewed against the
vendor's own documentation---the architecture manuals and intrinsics references
that define each instruction's
semantics~\cite{intel-sdm,intel-intrinsics-guide,arm-arm,arm-intrinsics}---to
ensure that our transcriptions and axioms faithfully capture the intended behavior.
We differentially test modelled operations against genuine
\rustinline{core::arch} intrinsics on feature-capable hardware,
providing empirical validation in addition to manual review.

Figure~\ref{tab:intrinsics} summarizes the intrinsics
we modelled, across two architectures and eight instruction-set extensions. 
These models are developed independently from the code 
that uses them to implement and verify algorithms. 
This limits the surface 
that needs to be reviewed and tested, and the risk of 
assuming what is needed in algorithm refinement proofs. 
We intend to upstream the resulting libraries to Aeneas.

\begin{figure*}[t]
\centering\small
\begin{tabular}{| l l l | r r r | r r r |}
\hline
\multicolumn{3}{|c|}{Component} & \multicolumn{3}{c|}{Specification} & \multicolumn{3}{c|}{Implementation} \\
Component & Standard & Status & Text & Core & Properties & Rust & Properties & Proofs \\
\hline
SHA-3 / SHAKE & FIPS 202    &                             &  1,847 & 140 & 2,276 & 1,741 &  3,732 & 10,823 \\
ML-KEM        & FIPS 203    &                             &  3,628 & 308 & 1,352 & 2,349 & 10,963 & 25,389 \\
AES / AES-GCM & FIPS 197 / SP 800-38D &                   &  6,978 & 205 & 2,020 & 3,194 &  9,760 & 21,815 \\
HMAC / HKDF   & RFC 2104/5869 & experimental              &  1,406 & 195 &   159 &   371 &    503 &  1,025 \\
SHA-2         & FIPS 180-4  & experimental                &  1,468 & 224 &   881 & 1,297 &  2,593 &  6,297 \\
ML-DSA        & FIPS 204    & experimental                &  4,192 & 500 & 4,076 & 3,352 & 12,427 & 35,006 \\
FrodoKEM      & draft-frodokem-02 & experimental          &  1,400 & 347 & 3,250 & 1,337 &  5,872 & 15,532 \\
HPKE          & RFC 9180    & experimental                &  4,810 & 502 &   278 & 1,082 &  3,934 &  6,116 \\
Intrinsics    & vendor manuals & verification-only        &      -- & 406 &   902 & 1,957 &  1,904 &  2,774 \\
\hline
Total         &             &                             & 25,729 & 2,827 & 15,194 & 16,680 & 51,688 & 124,777 \\
\hline
\end{tabular}
\caption{Verified SymCrypt Rust components; \emph{experimental} means the Rust source is not distributed as part of SymCrypt. 
Sizes are counted in \texttt{cloc} non-comment, non-blank source lines, 
except for Text, which counts lines of standard plaintext, extracted from PDF where applicable. 
In Specification, 
Core counts only the normative formalization of the standard (e.g., NTT),
or the trusted axioms for intrinsics; Properties counts the
additional material---properties, proofs, auxiliary definitions, and tests. 
In Implementation, 
Rust is source code excluding tests and 
Lean is split into
Properties: statements of correctness theorems, lemmas, 
and intermediate definitions; 
and Proofs: the tactic bodies that discharge them. 
In total, the formalization comprises
237,160 lines of Lean, of which 30,253 are Aeneas-generated Lean definitions
and 206,907 are source definitions and proofs.}
\label{tab:primitives}
\end{figure*}

\section{Case Study: Verified Rust in SymCrypt}
\label{sec:casestudy}

We present our main case study: verifying Rust code in SymCrypt, 
a widely-deployed cryptographic provider.
This code is subject to many design and implementation requirements 
that go beyond what is formally captured. For example, it must implement an existing C ABI, 
be compiled in \rustinline{no} \rustinline{std} mode, and 
share pre-existing internal data structures with C code.
It must also support a variety of platforms and compilers
(for deployment in Linux and in Windows, both in user and kernel mode), 
which imposes
tight bounds on its code size, its allocation discipline (with limits for heap and stack), 
and its handling of runtime errors. 
It must still support legacy platforms with no available Rust compilers. 
Pragmatically, this requires verifying Rust code as written in this context, 
with limited opportunity to change it to facilitate its verification, and 
accepting that some of its code (e.g. its C FFI) is outside the scope of verification. 

Figure~\ref{tab:primitives} lists the primitives we verified, their status, and
the size of the corresponding Rust code, specification, and proofs.
Formal specifications are much smaller than both the code and the standard prose, 
whereas the intermediate properties and proofs are much larger, but largely delegated to agents.
The number of primitives we verified allows us to cover
many patterns that were studied in previous work, with the difference that we tackle
them all at once in a single effort.

We kept the verification footprint on the SymCrypt source code minimal:
except for a \rustinline{lean} directory hosting all specifications and proofs 
(and ignored when compiling SymCrypt), 
we leverage the compile-time attribute \rustinline{"verify"} (see \sref{stdlib}) to
clearly mark the verification boundaries.
We use it, for example, to opacify code that calls unverifiable external functions,
to exclude extraction of FFIs and test-only code, and to switch between primitive and modelled intrinsics.

During verification, we discovered minor flaws and improvement opportunities in Rust code,
described below, suggesting that our methodology can provide 
not just higher assurance but also useful feedback to the programmer. 
The rest of this section discusses selected verification patterns and findings, grouped by families of algorithms. 

\subsection{SHA-3, SHAKE, and related algorithms} 

The SHA-3 family of hash functions illustrates `bit-scrambling' symmetric algorithms.
SymCrypt implements them in Rust, providing a 
stateful API for hashing multiple inputs  and extracting multiple outputs, 
used in turn, e.g., for pseudo-random sampling 
in post-quantum algorithms. 

\myparagraph{Keccak,} 
its core compression algorithm, 
repeatedly applies 5 elementary permutations 
to an internal 1600-bit state. It is inherently sequential;  
its performance hinges on instruction scheduling and avoiding state spilling. 

SymCrypt provides a portable textbook implementation and just relies on the compiler. 
We add two experimental variants: an unfolded fused loop 
that detects and (implicitly) uses BMI2's RORX on x86-64 for additional performance;
and an AVX2 variant of this loop that runs 4 separate instances of Keccak in parallel.
This SIMD code achieves an excellent trade-off between complexity and performance: 
a Rust macro explicitly provides the 3 intrinsics to use for 4-way rotations,
and the compiler efficiently autovectorizes the rest of the code.

Verifying Keccak is demanding because the unfolded loop bodies
perform hundreds of monadic computation steps on 25 U64s (or 4xU64s registers),
which need to be re-ordered and mapped back to the standard's bitwise specification. 
This is solved by combining Lean automation, Aeneas tactics, and nested applications of the \leaninline{decompose} command. 

\myparagraph{Incremental API and Ghost State.}
SymCrypt exposes SHA-3 and SHAKE through a stateful API
(\rustinline{new}, \rustinline{append}, \rustinline{result} for hashes, plus
\rustinline{extract} for extensible hash functions)
to incrementally feed input bytes and extract output bytes.
It also has a simpler one-shot API (\rustinline{hash} and \rustinline{shake}) 
that chains new, append, and result/extract together for a single input and output.
Meanwhile, the FIPS standard for SHA-3 is purely one-shot, with an
input/output equation defined once all input has been absorbed and all output has been squeezed.

SymCrypt maintains a runtime structure that holds the Keccak state 
as an array \rustinline{[u64; 25]} together 
with a flag that indicates whether it is currently absorbing or squeezing bytes,
the number of bytes processed so far, the number of usable bytes at the start of the
state between applications of the compression function, and a precomputed padding byte. 

To fully specify functional correctness, we supplement it with verification-only 
ghost state that records its byte history---the concatenations of all bytes \leaninline{absorbed} 
and \leaninline{squeezed} since initializing or resetting the state. 
Two predicates used in the pre- and post-conditions of the theorems associated with the methods relate ghost and runtime state: 
\leaninline{absorbing} asserts 
a given bytestring is being hashed, while \leaninline{squeezing} asserts
a given bytestring has been hashed (after padding and a final permutation) and a given bytestream 
has already been returned. 
See the resulting state machine in Fig.~\ref{fig:sha3-fsm}. 
Calling \rustinline{append} after \rustinline{result} or \rustinline{extract}
transparently resets the state. 

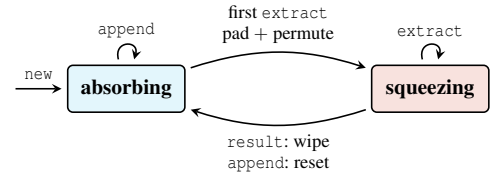
\begin{figure}[t]
\centering
\begin{tikzpicture}[>=stealth,semithick,
  mode/.style={draw,rounded corners=2pt,minimum width=1.55cm,
    minimum height=0.62cm,inner sep=2pt,font=\footnotesize\bfseries},
  action/.style={->,draw,font=\scriptsize,shorten >=2pt,shorten <=2pt}]
  \node[mode,fill=clAWSLC!18] (A) {absorbing};
  \node[mode,fill=clRust!14,right=2.45cm of A] (S) {squeezing};
  \draw[action] ([xshift=-0.75cm]A.west) -- node[above]{\texttt{new}} (A.west);
  \draw[action] (A) edge[loop above,looseness=5] node{\texttt{append}} (A);
  \draw[action] (S) edge[loop above,looseness=5] node{\texttt{extract}} (S);
  \draw[action,bend left=18] (A) to node[above,align=center]{first \texttt{extract}\\pad $+$ permute} (S);
  \draw[action,bend left=18] (S) to node[below,align=center]{\texttt{result}: wipe\\\texttt{append}: reset} (A);
\end{tikzpicture}
\vspace{-1ex}
\caption{SHA-3/SHAKE state machine. We supplement the runtime state with ghost state 
to precisely relate it to the one-shot FIPS-202 specification.}
\vspace{-1ex}
\label{fig:sha3-fsm}
\end{figure}

\myparagraph{Correctness Bug.}
In the process of verifying this state machine, we uncovered a 
bug in \rustinline{extract}. The routine drains a partial leading lane byte-by-byte,
copies whole 64-bit lanes, then writes a trailing partial lane; instead of 
advancing an output pointer between the three phases, the Rust port preserved
the cursor only for the byte loops and passed the \emph{entire} output slice to
the lane copy, which always writes from offset~0. When a call starts mid-lane
(\rustinline{state_index} not a multiple of~8), the lane block overwrites the
bytes just emitted and leaves uninitialized buffer content in the gap,
resulting in a few incorrect output bytes. 
The defect is dormant for one-shot calls and for \rustinline{result}, which always
enter \rustinline{extract} lane-aligned, but it is reachable from the multi-call
SHAKE128/256 XOFs that ML-KEM and ML-DSA use for sampling, leading to some 
wrong byte values in the output. 
The proof could not be closed along the misaligned branch; the proof agent automatically drafted a bug report
and suggested the one-line fix---slicing the output at the cursor before the lane copy.
(The SymCrypt team independently found this bug by differential testing.)

\myparagraph{Experimental Implementations}
We wrote and verified implementations of SHA-2, 
relying on intrinsics for its 256-bit compression function and 
re-using of the SHA-3 verification patterns, as well as HMAC and HKDF
(used for key derivation in TLS and HPKE), illustrating the verification of Rust traits for algorithms parameterized by hash algorithms.

\subsection{ML-KEM}

ML-KEM is the main standard for post-quantum key encapsulation.
Based on error correction, it relies on complex algorithms involving both maths 
and platform-specific optimizations.
SymCrypt implements it in Rust, as a stateful API to amortize pre-computations on keys. 
Appendix~\ref{sec:mlkem-api} presents the theorem statement
for its decapsulation function. 

Besides termination, discussed below, the main verification challenges are 
in its NTT implementation, using modulo arithmetic on vectors and matrices, 
specialized with intrinsics for 4 different architectures and feature sets to verify together;
in compressing followed by encoding (and the reverse), concatenating elements into 
an array of bits, which requires an invariant akin to the one for streaming hash
implementations~\cite{ho2023hacl};
the efficient sampling based on SHAKE; and the top-level API, which involves a bulky state
invariant on internal pre-computed key materials. 

\myparagraph{Minor Code Defects.} 
We identified an arithmetic bug in a runtime assertion claiming $x < Q$ 
for an intermediate coefficient in the NTT. The proof agent produced a counterexample leading to $x = Q$,
then completed the proof for the fixed statement $x \leq Q$. 
This bug was independently found by external audit. 
We also found and fixed erroneous claims in comments informally reasoning about similar
arithmetic bounds.

In the other direction, we leveraged functional verification to tighten 
some bounds and simplify the code, for example eliminating unnecessary modulo reductions,
and loops in \rustinline{compress_and_encode} and \rustinline{decode_and_decompress} after proving that a single iteration always suffices.

\myparagraph{Sampling Termination} 
ML-KEM internally relies on rejection sampling to uniformly pseudo-randomly fill large arrays of coefficients in $\mathbb{Z}/q\mathbb{Z}$ where $q = 3329$ (FIPS 203, Algorithm 7). 
The sampling loop repeatedly consumes the output of a SHAKE128 extensible hash function (XOF)
seeded with key material and indices, 3 bytes at a time, to produce two 12-bit candidates (since $3329 < 2^{12}$), and it accepts any candidate smaller than $q$. This process continues until the array is filled, with no obvious bound on the number of SHAKE128 output bytes it consumes.
Aeneas allows proving partial correctness, that is: sampling either loops or
terminates on an output satisfying its post-condition; the downside is that
the verification of all the upper layers would need to be stated as a \emph{partial}
correctness result.
Instead, we prove termination
by proving (as a specification-level property of SHA-3) that
the Keccak function used to define the XOF is a permutation on its 1600-bit state space and 
by formalizing the elegant argument of Barbosa and Schwabe~\cite{barbosa2023kyber}: 
the streams of output bytes produced by repeatedly applying Keccak 
are periodic (for some astronomically large period) and include its pre-image;
this pre-image 
is the 1600-bit initial state of the XOF,  
which includes 43 byte-aligned zero triples that will eventually
provide enough candidates smaller than $q$ to fill the array after a few periods. 
Equipped with the theorem stating the existence of a termination bound,  
we can resume Rust verification and prove that the implementation function also terminates. 
(The 1,400-line proof was automatically developed by an agent in less than an hour.) 

\subsection{AES and AES-GCM} SymCrypt is migrating AES and AES-GCM to Rust, 
with drop-in replacements for key generation, core AES,
and optimized platform-specific loop bodies for AES-GCM provided as a partial implementation 
of the existing C ABI. We also provide experimental code to stitch this code into full-fledged verified Rust implementations of the AES and AES-GCM standards. 
This code includes large functions and macros consisting mostly of intrinsics for AES and GCM, 
pushing the limits of proof complexity for Lean automation. 

\subsection{FrodoKEM}

FrodoKEM~\cite{frodokem} is an alternative to ML-KEM with a simpler, more conservative design. 
We first verified its existing reference Rust implementation, then developed and verified an experimental variant integrated with SymCrypt: with the same stateful API as ML-KEM, improved performance, and reduced code size and memory footprint. 

The main performance and verification bottleneck consists of 
two variants of matrix multiplication involving a large pseudo-random matrix (sampled one strip at a time), each tiled with 6 carefully-written loops to yield good implicit autovectorization for our target architectures. (We confirmed using intrinsics would not significantly improve them.) 
Verification also involves reasoning about bitvectors and signed and unsigned integers,
parametrically in their width, and verified integration with AES and SHAKE-128/256 for sampling.

\myparagraph{Correctness Bug} We found a bug in the constant-time comparison 
of its Fujisaki-Okamoto transform.
(This bug dangerously results in returning a value that depends on the decapsulated secret even
when re-encryption fails.)
We list below the Rust function, where \texttt{>{}> 8} should be \texttt{>{}> 16}
\begin{minted}{rust}
// Constant-time equality mask
// Returns 0xFF if the arrays are equal, else 0x00
fn ct_equal_u16(a: &[u16], b: &[u16]) -> u8 {
  debug_assert_eq!(a.len(), b.len());
  let mut diff: u16 = 0;
  for i in 0..a.len() {
    diff |= a[i] ^ b[i];
  }
  ((diff as u32).wrapping_sub(1) >> 8) as u8
}
\end{minted}
and the corresponding theorem statement, 
stating that this function correctly implements comparison between two arrays of u16s of the same length.
\begin{minted}{lean}
theorem ct_equal_u16.spec  (a b : Slice Std.U16)
  (h_len : a.length = b.length) : frodokem.ct_equal_u16 a b
    ⦃ m => m = (if a = b then 0xFF#u8 else 0x00#u8) ⦄ 
\end{minted}
The function is short but complicated by constant-time encodings
and coercions between u8, u16, and u32. 
Once patched, it is automatically verified by Aeneas tactics. 

Verification also led to minor simplifications, removing all intermediate normalization steps and an unnecessary iteration in sampling based on constant arrays of CDF coefficients. 

\subsection{ML-DSA}
ML-DSA is the main standard for post-quantum signature; its design 
is related to ML-KEM but more complex (the standard defines 49 algorithms). 
SymCrypt provides a portable C implementation, 
so we jointly produced and verified an experimental Rust implementation, 
coded by an agent instructed to follow the RFC (for
correctness),
the existing C code (for API and low-level choices), and the  
related Rust ML-KEM code (for SymCrypt integration). 
We then refined it with various optimizations, notably using our verified 4-way SHAKE 
implementation to accelerate sampling. 
Its verified code passes the SymCrypt tests and yields major performance gains. 

Verification highlights include more NTTs and the generalization and re-use of ML-KEM's termination theorem 
on three sampling functions. ML-DSA's signing algorithm may independently fail to terminate
(with negligible probability), so we prove that its implementation, 
which is parameterized by a maximal number of pseudo-random attempts
(as in the C code) succeeds on valid arguments iff any of these attempts succeeds at the specification level.

\subsection{HPKE}
We finally implemented HPKE, to confirm that we could verifiably compose our algorithm implementations 
(ML-KEM, HKDF, HMAC, SHA-3/SHAKE, SHA-2, AES-GCM, AES) into 
full-fledged post-quantum-secure ciphersuites. 
The main verification challenge is to handle its modular design and cryptographic agility.

\section{Evaluation}
\label{sec:evaluation}

\subsection{Agentic Proofs}
\label{sec:eval-ai}

As in the field of mathematics, where AI models are now able to prove or disprove open problems
and conjectures backed by Lean formalization (e.g.~\cite{openai-lean}),
we find that frontier models perform well at program verification tasks in Lean, 
which provides clear goals to
work towards and detects most hallucinations and errors. 
Models still try to ``cheat'' with unsound assumptions
or weakened goals, so careful human review of theorem statements remains essential.

We used GitHub Copilot with different agents in the course of the project, starting with Claude Opus
4.6 set to high reasoning and 264K tokens of context window, which completed most proofs.
Some hard proofs, like FrodoKEM's encapsulation, required a larger context window (1M).
We then switched to Opus 4.7, 4.8, GPT 5.6 Sol, and finally Opus 5.0.
Overall, agents generated $\sim$5.1k theorems for a total of 125 KLOC of proofs,
excluding theorem statements and auxiliary definitions.
We estimate the average verification cost from scratch to be
a few dollars per line of Rust code (at the time of writing),
with high variance by algorithm and model used.
The generated proof scripts are more verbose than human-written proofs,
but we kept them tractable through Lean automation, skill files, and prescriptive prompts;
more than half the proof volume accounts for theorems with less than 100 lines of proofs,
while only 14 proofs span more than 500 lines, with a maximum at 1.1 KLOC.

We evaluate the time and cost of AI-generated proofs on 7 hard proofs
from ML-KEM,
HKDF and SHA-3 that we estimate would each require hours and hundreds of lines of Lean to write by hand.
Figure~\ref{fig:agentic-cost} compares Opus 4.8, GPT 5.6 Sol and Opus 5
on this suite, in terms of wall time, success rate, proof lines and dollar cost.
Each bar averages the successful terminal runs of the suite; cost is billed inference usage converted to US dollars.
Opus 5 dominates: it is the only model that closes all 7 proofs within single sessions, and it does
so about $2.8\times$ faster and $2.5\times$ cheaper than Opus 4.8, for a comparable proof length. 
This indicates a rapid improvement trend in frontier model capabilities for our use case. 

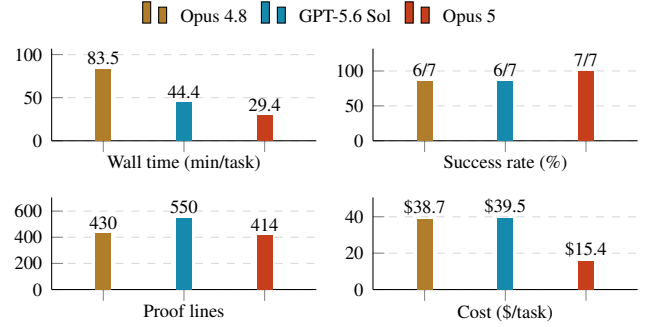
\begin{figure}[t]
\centering
\pgfplotsset{
  agenticbar/.style={
    ybar,
    width=0.6\columnwidth, height=2.8cm,
    bar width=6pt,
    ymin=0,
    enlarge x limits=0.32,
    enlarge y limits={upper,value=0.28},
    symbolic x coords={O48,Sol,O5},
    xmin=O48, xmax=O5,
    xticklabels={,,},
    every axis plot/.append style={bar shift=0pt, draw=none},
    nodes near coords, point meta=explicit symbolic,
    every node near coord/.append style={font=\scriptsize, inner sep=1.5pt},
    tick label style={font=\scriptsize},
    xlabel style={font=\scriptsize, yshift=9pt},
    ylabel style={font=\scriptsize},
    ymajorgrids, grid style={dashed, gray!30},
    axis lines*=left,
    legend style={font=\scriptsize, draw=none, fill=none, legend columns=-1,
      column sep=3pt},
  }
}
\pgfplotslegendfromname{agenticlegend}

\begin{tikzpicture}[baseline]
  \begin{groupplot}[agenticbar,
    group style={group size=2 by 2, horizontal sep=0.75cm, vertical sep=0.75cm}]

  \nextgroupplot[xlabel={Wall time (min/task)}, legend to name=agenticlegend]
    \addplot[fill=clOpus48] coordinates {(O48,83.5) [83.5]};
    \addplot[fill=clSol]    coordinates {(Sol,44.4) [44.4]};
    \addplot[fill=clOpus5]  coordinates {(O5,29.4)  [29.4]};
    \legend{Opus 4.8, GPT-5.6 Sol, Opus 5}
  \nextgroupplot[xlabel={Success rate (\%)}, ymax=100, enlarge y limits={upper,value=0.32}]
    \addplot[fill=clOpus48] coordinates {(O48,85.7) [6/7]};
    \addplot[fill=clSol]    coordinates {(Sol,85.7) [6/7]};
    \addplot[fill=clOpus5]  coordinates {(O5,100)   [7/7]};
  \nextgroupplot[xlabel={Proof lines}]
    \addplot[fill=clOpus48] coordinates {(O48,429.8) [430]};
    \addplot[fill=clSol]    coordinates {(Sol,550.3) [550]};
    \addplot[fill=clOpus5]  coordinates {(O5,413.9)  [414]};
  \nextgroupplot[xlabel={Cost (\$/task)}]
    \addplot[fill=clOpus48] coordinates {(O48,38.75) [\$38.7]};
    \addplot[fill=clSol]    coordinates {(Sol,39.54) [\$39.5]};
    \addplot[fill=clOpus5]  coordinates {(O5,15.41)  [\$15.4]};
  \end{groupplot}
\end{tikzpicture}

\caption{Cost of agentic proofs, comparing Opus 4.8, GPT-5.6 Sol and Opus 5 on the hard-proof
suite of 7 proofs from HKDF, ML-KEM and SHA-3. Wall time, proof lines and cost are means over the successful terminal runs of the
suite.}
\label{fig:agentic-cost}
\end{figure}

\subsection{Integration with SymCrypt}

Moving from C to Rust adds new dependencies to its compiler and toolchain, incurring costs
in terms of integration, build, tooling and CI. 
For instance, the Rust approach of linking all dependencies into the final executable
might cause issues, either for provenance (distributing a security-critical
update for a library requires recompiling every client) or compliance (the
FIPS cryptographic standard is designed for a shared library exposing a stable
C-like ABI).
Another challenge is the ability to use link-time and runtime optimization based
on the platform and features of the CPU. Generally, Rust requires the code to be compiled
for the target architecture to be fully optimized, whereas SymCrypt expects to link the same binary
for modern 64-bit with AVX512 or ancient 32-bit x86.

To mitigate some of these issues while migrating, we
developed a compiler from Rust to C for the sake of backwards compatibility, dubbed
Eurydice.
Eurydice is built on top of the C-producing KaRaMeL compiler~\cite{lowstar}.
It consumes the LLBC emitted by Charon and relies on a pipeline of nano-passes to produce
C code.
After careful evaluation, the SymCrypt team plans to ship all compiled Rust code directly,
despite the optimization loss of statically linking a binary not optimized for the target platform,
including verified SHA-3 and ML-KEM. This approach preserves FIPS certification.

\subsection{Performance}
\label{sec:performance}

We compare the performance of the verified Rust implementation
against the original SymCrypt~C code (our baseline, normalized to~100),
OpenSSL~4.0.1 (to compare against other highly optimized C implementations),
and the other verified crypto libraries that have seen wide deployment:
\hacl~\cite{zinzindohoue2017hacl} (verified in \fstar, used in Firefox and WireGuard),
Formosa~\cite{jasmin} (written in Jasmin, used in Signal and PQShield),
AWS-LC~\cite{amazoncrypto} (verified with HOL-Light, Isabelle, SAW, CBMC,
used in many Amazon services), and libcrux~\cite{libcrux} (verified with
hax~\cite{bhargavan2024hax} and used in Firefox, Signal and OpenSSH).

All benchmarks were collected on the same x86-64 platform (Intel Xeon
W-2155 Skylake-SP, AES-NI and AVX2 enabled); throughput is measured on
16\,KB messages (asymptotic regime); latency is the median over at least
500 iterations.
Figure~\ref{fig:perf-sym} shows relative \emph{throughput} for
symmetric primitives (higher is better);
Figure~\ref{fig:perf-pq}
shows relative \emph{speed} for post-quantum operations and for HPKE
(higher is better).  In all figures the dashed line marks the baseline,
which is SymCrypt~C except for the experimental FrodoKEM and HPKE.

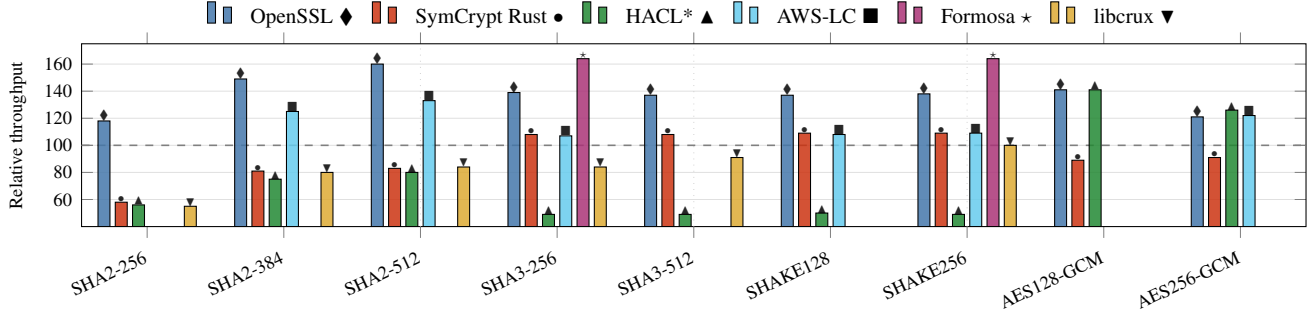
\begin{figure*}[t]
\centering
\begin{tikzpicture}
\begin{axis}[perfbar,
  width=\textwidth, height=4cm,
  symbolic x coords={SHA2-256,SHA2-384,SHA2-512,SHA3-256,SHA3-512,
                     SHAKE128,SHAKE256,AES128-GCM,AES256-GCM},
  xtick=data,
  xticklabel style={rotate=25, anchor=north east},
  ymin=40, ymax=175,
  ytick={60,80,100,120,140,160},
  ylabel={Relative throughput},
  extra y ticks={100},
  extra y tick labels={},
  extra y tick style={grid=major,
    major grid style={dashed,black!50,line width=0.6pt}},
  bar width=4.5pt,
  enlarge x limits=0.06,
  legend style={font=\footnotesize, at={(0.5,1.03)}, anchor=south,
    legend columns=6, column sep=5pt, draw=none, fill=none},
  extra x ticks={SHA2-512,SHA3-512,SHAKE256},
  extra x tick labels={},
  extra x tick style={grid=major,
    major grid style={dotted,black!20,line width=0.4pt},
    tick style={draw=none}},
]
\addplot[fill=clOSSL,
  nodes near coords={$\blacklozenge$},
  nodes near coords style={font=\tiny, anchor=south, inner sep=0pt}]
  coordinates {
    (SHA2-256,118) (SHA2-384,149) (SHA2-512,160)
    (SHA3-256,139) (SHA3-512,137)
    (SHAKE128,137) (SHAKE256,138)
    (AES128-GCM,141) (AES256-GCM,121)};
\addplot[fill=clRust,
  nodes near coords={$\bullet$},
  nodes near coords style={font=\tiny, anchor=south, inner sep=0pt}]
  coordinates {
    (SHA2-256,58)  (SHA2-384,81)  (SHA2-512,83)
    (SHA3-256,108) (SHA3-512,108)
    (SHAKE128,109) (SHAKE256,109)
    (AES128-GCM,89) (AES256-GCM,91)};
\addplot[fill=clHACL,
  nodes near coords={$\blacktriangle$},
  nodes near coords style={font=\tiny, anchor=south, inner sep=0pt}]
  coordinates {
    (SHA2-256,56)  (SHA2-384,75)  (SHA2-512,80)
    (SHA3-256,49)  (SHA3-512,49)
    (SHAKE128,50)  (SHAKE256,49)
    (AES128-GCM,141) (AES256-GCM,126)};
\addplot[fill=clAWSLC,
  nodes near coords={$\blacksquare$},
  nodes near coords style={font=\tiny, anchor=south, inner sep=0pt}]
  coordinates {
    (SHA2-256,nan) (SHA2-384,125) (SHA2-512,133)
    (SHA3-256,107) (SHA3-512,nan)
    (SHAKE128,108) (SHAKE256,109)
    (AES128-GCM,nan) (AES256-GCM,122)};
\addplot[fill=clFormosa,
  nodes near coords={$\star$},
  nodes near coords style={font=\tiny, anchor=south, inner sep=0pt}]
  coordinates {
    (SHA2-256,nan) (SHA2-384,nan) (SHA2-512,nan)
    (SHA3-256,164) (SHA3-512,nan)
    (SHAKE128,nan) (SHAKE256,164)
    (AES128-GCM,nan) (AES256-GCM,nan)};
\addplot[fill=clCrux,
  nodes near coords={$\blacktriangledown$},
  nodes near coords style={font=\tiny, anchor=south, inner sep=0pt}]
  coordinates {
    (SHA2-256,55)  (SHA2-384,80)  (SHA2-512,84)
    (SHA3-256,84)  (SHA3-512,91)
    (SHAKE128,nan) (SHAKE256,100)
    (AES128-GCM,nan) (AES256-GCM,nan)};

\legend{OpenSSL\;$\blacklozenge$,
        SymCrypt Rust\;$\bullet$,
        HACL*\;$\blacktriangle$,
        AWS-LC\;$\blacksquare$,
        Formosa\;$\star$,
        libcrux\;$\blacktriangledown$}
\end{axis}
\end{tikzpicture}
\vspace{-2mm}
\caption{Symmetric-crypto throughput on 16\,KB messages, relative to
  SymCrypt~C ($=100$, dashed line), reporting the median over 500 messages.
  Missing bars indicate unsupported algorithms.}
\label{fig:perf-sym}
\end{figure*}

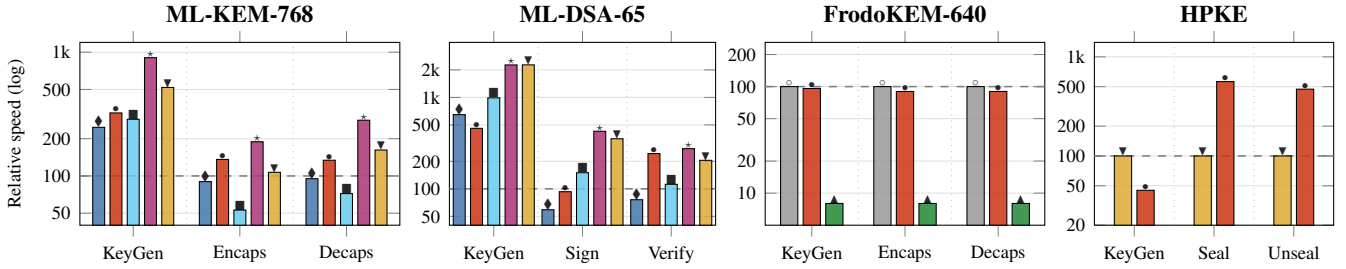
\begin{figure*}[t]
\centering
\pgfplotsset{
  pqpanel/.style={
    perfbar,
    height=4cm,
    enlarge x limits=false,
    xmin=0.5, xmax=3.5,
    xtick={1,2,3},
    xticklabel style={font=\scriptsize},
    ylabel={},
    extra y ticks={100},
    extra y tick labels={},
    extra y tick style={grid=major,
      major grid style={dashed,black!50,line width=0.6pt}},
    extra x ticks={1.5,2.5},
    extra x tick labels={},
    extra x tick style={grid=major,
      major grid style={dotted,black!30,line width=0.4pt},
      tick style={draw=none}},
  },
  pqmark/.style={nodes near coords style={font=\tiny, anchor=south, inner sep=0pt}},
}
\begin{tabular}{@{}c@{\hspace{1pt}}c@{\hspace{1pt}}c@{\hspace{1pt}}c@{}}
\begin{tikzpicture}
\begin{axis}[pqpanel,
  ymode=log, log origin=infty,
  width=0.327\textwidth,
  xticklabels={KeyGen,Encaps,Decaps},
  title={\small ML-KEM-768},
  ymin=40, ymax=1200,
  ytick={50,100,200,500,1000},
  yticklabels={50,100,200,500,1k},
  ylabel={Relative speed (log)},
  bar width=4.5pt,
]
\addplot[fill=clOSSL, pqmark, nodes near coords={$\blacklozenge$}]
  coordinates {(1,247) (2,90)  (3,95)};
\addplot[fill=clRust, pqmark, nodes near coords={$\bullet$}]
  coordinates {(1,323) (2,136) (3,134)};
\addplot[fill=clAWSLC, pqmark, nodes near coords={$\blacksquare$}]
  coordinates {(1,287) (2,53)  (3,72)};
\addplot[fill=clFormosa, pqmark, nodes near coords={$\star$}]
  coordinates {(1,903) (2,189) (3,282)};
\addplot[fill=clCrux, pqmark, nodes near coords={$\blacktriangledown$}]
  coordinates {(1,519) (2,107) (3,162)};
\end{axis}
\end{tikzpicture}
&
\begin{tikzpicture}
\begin{axis}[pqpanel,
  ymode=log, log origin=infty,
  width=0.287\textwidth,
  xticklabels={KeyGen,Sign,Verify},
  title={\small ML-DSA-65},
  ymin=40, ymax=4000,
  ytick={50,100,200,500,1000,2000},
  yticklabels={50,100,200,500,1k,2k},
  bar width=4.5pt,
]
\addplot[fill=clOSSL, pqmark, nodes near coords={$\blacklozenge$}]
  coordinates {(1,648)  (2,59)  (3,76)};
\addplot[fill=clRust, pqmark, nodes near coords={$\bullet$}]
  coordinates {(1,459)  (2,93)  (3,243)};
\addplot[fill=clAWSLC, pqmark, nodes near coords={$\blacksquare$}]
  coordinates {(1,989)  (2,150) (3,112)};
\addplot[fill=clFormosa, pqmark, nodes near coords={$\star$}]
  coordinates {(1,2266) (2,426) (3,275)};
\addplot[fill=clCrux, pqmark, nodes near coords={$\blacktriangledown$}]
  coordinates {(1,2270) (2,353) (3,205)};
\end{axis}
\end{tikzpicture}
&
\begin{tikzpicture}
\begin{axis}[pqpanel,
  ymode=log, log origin=infty,
  width=0.297\textwidth,
  xticklabels={KeyGen,Encaps,Decaps},
  title={\small FrodoKEM-640},
  ymin=5, ymax=260,
  ytick={10,20,50,100,200},
  yticklabels={10,20,50,100,200},
  bar width=6.5pt,
]
\addplot[fill=clLweke, pqmark, nodes near coords={$\circ$}]
  coordinates {(1,100) (2,100) (3,100)};
\addplot[fill=clRust, pqmark, nodes near coords={$\bullet$}]
  coordinates {(1,96)  (2,90)  (3,90)};
\addplot[fill=clHACL, pqmark, nodes near coords={$\blacktriangle$}]
  coordinates {(1,8)   (2,8)   (3,8)};
\end{axis}
\end{tikzpicture}
&
\begin{tikzpicture}
\begin{axis}[pqpanel,
  ymode=log, log origin=infty,
  width=0.267\textwidth,
  xticklabels={KeyGen,Seal,Unseal},
  title={\small HPKE},
  ymin=20, ymax=1400,
  ytick={20,50,100,200,500,1000},
  yticklabels={20,50,100,200,500,1k},
  bar width=6.5pt,
]
\addplot[fill=clCrux, pqmark, nodes near coords={$\blacktriangledown$}]
  coordinates {(1,100) (2,100) (3,100)};
\addplot[fill=clRust, pqmark, nodes near coords={$\bullet$}]
  coordinates {(1,45)  (2,563) (3,472)};
\end{axis}
\end{tikzpicture}
\end{tabular}
\vspace{-1mm}
\caption{Post-quantum and hybrid-encryption latency, on a log scale,
  relative to SymCrypt~C ($=100$, dashed line) for ML-KEM-768 and ML-DSA-65,
  to the lweke reference~C for FrodoKEM-640, and to libcrux for HPKE (the latter
  2 are not supported in SymCrypt). Colors and markers match Figure~\ref{fig:perf-sym}.
  operations.}
\label{fig:perf-pq}
\end{figure*}

\myparagraph{Symmetric primitives}
For SHA-3/SHAKE, the Rust compiler auto-vectorizes the Keccak permutation of SHA-3 
(which is used by ML-KEM and ML-DSA) effectively: SymCrypt's Rust code
is 8\% \emph{faster} than the C code, about 15\% faster than libcrux, and matches AWS-LC,
with HACL* lagging by over 40\%.
For SHA-2, SymCrypt Rust is ${\sim}40$\% slower than SymCrypt C, which uses
SIMD-optimized compression rounds; this gap narrows to ${\sim}20$\%
for the 64-bit variants (SHA-384/512). This is similar to libcrux
and HACL*, but worse than AWS-LC which has been manually vectorized.
For AES-GCM, the stitched Rust kernel reaches ${\sim}89$\% of
SymCrypt's SSE2 throughput (without vectorized carry-less multiplication)
OpenSSL and HACL* exceed SymCrypt by 20--40\% thanks to pipelining.
libcrux's AES-NI path is $6{-}12\times$ slower, due to
suboptimal GHASH interleaving.

\myparagraph{Post-quantum primitives}
For ML-KEM, SymCrypt Rust with AVX2 intrinsics for the NTT is
$1.3{-}1.4\times$ faster than SymCrypt C for encapsulation
and decapsulation across all three parameter sizes, and competitive
with OpenSSL.
Formosa's Jasmin-generated AVX2 assembly is the fastest
verified implementation at $2{-}3\times$ for encaps/decaps
and ${\sim}9\times$ for KeyGen.
For ML-DSA, SymCrypt Rust matches SymCrypt C's signing speed
and achieves $2.3{-}2.5\times$ faster verification across all sizes,
owing to vectorization and the more optimized Keccak.
Formosa AVX2 leads at $4{-}5\times$ for signing and
${\sim}3\times$ for verification.
FrodoKEM (not implemented in SymCrypt C) is benchmarked against the
reference C implementation.  SymCrypt Rust is within 10\% of the reference;
HACL*'s verified implementation is ${\sim}12\times$ slower.

\myparagraph{Hybrid public-key encryption}
We compare our verified implementation of HPKE with libcrux (via the {\tt hpke-rs} crate),
using the ML-KEM-768 / HKDF-SHA256 / AES-128-GCM suite. SymCrypt doesn't offer HPKE
via its C API yet. While HACL* supports HPKE, it doesn't offer post-quantum KEM, so is not included.
On 16\,KB payloads, SymCrypt Rust seals in $17.8\,\mu s$ and unseals in
$19.3\,\mu s$, against $100.2\,\mu s$ and $90.8\,\mu s$ for libcrux, i.e.
$5.6\times$ and $4.7\times$ faster, though key generation from ML-KEM is twice slower.

\section{Related Work}
\label{sec:related}

We discuss two main lines of related work aiming at verifying implementations 
of cryptographic algorithms.

\myparagraph{Verification-Oriented Code.}
Several prior works develop implementations 
in verification-oriented languages.
They often write generic code 
that can be verified once, then specialized for different targets and parameters,
and compiled or extracted to C or assembly. 

The Jasmin language~\cite{jasmin} combines high-level features, such as procedures and loops,
with assembly-level instructions, and ships using  a compiler
certified in Rocq; Jasmin programs are also translated to EasyCrypt to verify
functional correctness, cryptographic security, and timing-attack
resistance~\cite{easycryptSha3,jasmin23kyber,jasmin24kyber}.
Jasmin implementations are platform-specific and allow tight control over the choice of
registers and assembly instructions.

FiatCrypto~\cite{fiat-crypto} uses Rocq to verify modular arithmetic libraries parametric in 
the representation of elements in a prime field and the underlying hardware architecture.

The Vale language~\cite{vale,vale-fstar} allows writing assembly-like
implementations parametric in the underlying assembler and operating system's
calling conventions.

\hacl~\cite{haclxn} uses \lowstar~\cite{lowstar}, a C-like language embedded in \fstar~\cite{fstar},
to implement and verify code specialized for multiple backends,
including all the primitives needed for a TLS cipher suite~\cite{zinzindohoue2017hacl}
and a streaming functor for hashing algorithms~\cite{ho2023hacl}.

Deployment, e.g., to Mozilla's NSS or Python for \hacl, and Google's BoringSSL for FiatCrypto, 
involves translating the verified code to mainstream languages like C or assembly.
Despite significant effort to generate idiomatic, maintainable code, this hinders 
integration into existing codebases. For instance, adopters of the EverCrypt cryptographic provider~\cite{evercrypt} commonly kept their own unverified infrastructure for multiplexing
and feature detection~\cite{everestperspectives}.
In contrast, we target Rust code written by software engineers, 
designed to be portable while supporting target-specific implementations, 
and already integrated into their target platforms and products.

\myparagraph{Implementations in Mainstream Languages.}
Other works investigate the direct verification of existing implementations.
Several projects rely on the Verified Software Toolchain~\cite{vst}
to verify C implementations of different primitives deployed in
OpenSSL~\cite{appel-hmac,appel2015sha} and mbedTLS~\cite{appel-hmac-drbg}.

CryptoLine uses a combination of proof assistants, automated SMT solvers, and computer
algebra systems to reason about handwritten assembly implementations of
field and group
operations~\cite{tsai17crypto,chen2014curve25519,polyakov2018assembly}
and NTT multiplications~\cite{cryptolineNTT}.
Dockins et al.~\cite{galois-saw} use symbolic execution with SAW to establish the equivalence of C and Java implementations with respect
to reference implementations written in Cryptol~\cite{cryptol}.

Apple~\cite{applecrypto} combines SAW with a translation from Cryptol to Isabelle,
proving equivalence between C code, Cryptol models, and Isabelle formalizations of the standards.  
They verify portable C implementations of ML-KEM and ML-DSA, and prove 
their ARM64 subroutines equivalent to the C code they replace. 
One stated limitation is that SAW does not support arbitrary message lengths in ML-DSA,
which are instead thoroughly tested. 

Becker et al. combined HOL Light~\cite{harrison2009hol} and CBMC~\cite{kroening2014cbmc}
to verify an implementation of ML-KEM integrated into Amazon's AWS-LC~\cite{amazoncrypto}.
They establish functional correctness for core components written in assembly,
as well as memory safety and the absence of integer overflows for code written in C (via bounded model checking).

Libcrux~\cite{libcrux} uses hax~\cite{bhargavan2024hax}
to verify the functional correctness of Rust implementations for SHA-3, ML-KEM, and parts of ML-DSA. 
Specifications and proofs are embedded in Rust using attributes, then
extracted with the code to pure models in~\fstar.

In contrast, we clearly separate code and proofs, 
enabling the deployment of verified code with minimal annotations, 
and the verification of high-level properties using 
Lean's rich mathematical libraries. 
Our focus on Rust and Lean enables us to efficiently model and verify 
functional correctness and panic freedom for entire high-performance implementations
of algorithms integrated into SymCrypt.

\section*{Acknowledgements}

We thank the SymCrypt team, in particular Samuel Lee, Jason Fisher, and Jessica Krynitsky,
for their close collaboration and support throughout this project.
We also thank Mantas Baksis and Karthik Bhargavan for valuable discussions and feedback.
Joshua Clune's contributions were made while he was an intern at Microsoft Research.
Michael Naehrig's contributions were made while he was at Microsoft.
Work at Inria received funding from a France 2030 program managed by the French National
Research Agency under grant agreement ANR-22-PETQ-0008 PQ-TLS.

\section*{Public Release}

We have released a significant subset of the formal development described in this draft,
including the verified SymCrypt code, formal specifications, and proofs for ML-KEM and SHA-3,
at \url{https://github.com/microsoft/SymCrypt/tree/feature/verifiedcrypto}.
We plan to release further parts of the development as they mature.
An overview of the project is available on the Microsoft Research blog at
\url{https://www.microsoft.com/en-us/research/blog/verifying-rust-cryptography-in-symcrypt-from-standards-to-code/}.
This draft will be superseded by a forthcoming technical report.

\appendix

\section{Multi-Target Verification Example}\label{sec:nttarch}

Consider the Rust code snippet below, simplified from ML-KEM,
showing support for two different NTT back-ends for x86 and AArch64.

\begin{minted}{rust}
fn ntt_layer(src: &mut Element, k: usize, len: usize) {
  #[cfg(any(target_arch = "x86", 
            target_arch = "x86_64"))]
  { ntt_layer_x86(src, k, len); }
  #[cfg(target_arch = "aarch64")]
  { ntt_layer_aarch64(src, k, len); } }
\end{minted}

Given a \emph{set} of targets that include those mentioned in this code, 
Charon and Aeneas extract it to a Lean model that turns conditional compilation into dynamic branching:

\begin{minted}{lean}
def ntt_layer src k len : Result Element := do
  let tgt ← get_target
  if tgt = "x86-64" || tgt = "x86" 
  then ntt_layer_x86 src k len
  else ntt_layer_aarch64 src k len
\end{minted}
For readability, the target names above abbreviate the full target triples emitted by Aeneas.
For verification, \leaninline{get_target} is underspecified;
hence this code requires verifying all branches against the standard.

\section{Intrinsics Examples}\label{sec:intrinsics-examples}

We illustrate the two kinds of shims for intrinsics discussed in \sref{intrinsics},
starting with Rust models.
The AVX2
\rustinline{_mm256_set1_epi16} intrinsic broadcasts one 16-bit value to all lanes.
Using an unsigned lane view, it can be modelled as the Rust function:
\begin{minted}{rust}
fn set1_epi16(a: u16) -> [u16; 16] { 
  assert!(cpu_features_present(FEATURE_AVX2));
  [a; 16] }
\end{minted}
translated into Lean and verified against an equivalent pure specification,
using a theorem of the form:
\begin{minted}{lean}
theorem set1_epi16.spec (a : U16) 
  (hfeat : featurePresent FEATURE_AVX2 = true) :
  set1_epi16 a ⦃ r =>
    ∀ k (hk : k < 16), r[k] = a ⦄ := by ...
\end{minted}
The assertion is translated as a call to \leaninline{massert}; hence proving
this theorem, or using it at a call site, requires \leaninline{hfeat}.

In contrast, an intrinsic that applies one round of AES 
to a register given the round key in another register 
is modelled by an axiom referring to our formalization of the AES standard, of the form:
\begin{minted}{lean}
axiom mm_aesenc_si128.spec r rkey :
  mm_aesenc_si128 r rkey ⦃ r' =>
    r'.toBytes = Spec.AES.round r.toBytes rkey.toBytes ⦄
\end{minted}

\section{Verified ML-KEM API Example}
\label{sec:mlkem-api}

We illustrate the top-level theorems we prove for the main functions that form the public interface of our cryptographic implementations using the decapsulation function of ML-KEM as a concrete example.   
The Rust function has signature
\begin{minted}{rust}
pub fn decapsulate(
    pk_mlkem_key: &Key,
    pb_ciphertext: &[u8],
    pb_agreed_secret: &mut [u8],
) -> Error
\end{minted}
where \rustinline{pk_mlkem_key} is a key object, previously created by either generating or loading a private key, \rustinline{pb_ciphertext} is the input buffer holding the ML-KEM ciphertext, and \rustinline{pb_agreed_secret} is the output buffer for the decapsulated secret. 
Note that decapsulation can return an error. 

We list below the corresponding Lean theorem statement, where
\leaninline{mlkem.decapsulate} is the name of the Lean function extracted from \rustinline{decapsulate}.
The theorem takes two additional parameters: 
\leaninline{params}, the algorithmic parameters of ML-KEM (controlling strength, key lengths, etc)
provided when creating the key object, and \leaninline{h_key}, a precondition using an abstract state predicate \leaninline{wfDecapKey}, which can be established 
as a post-condition of key creation using these parameters. 
This predicate records structural well-formedness, 
and the presence and consistency of the cached private key material with FIPS~203. 
Importantly for standard compliance, it also records that the mandatory decapsulation-key validation was performed during generation or import.

\begin{minted}{lean}
theorem mlkem.decapsulate.spec
    (params : ParameterSet)
    (pk_mlkem_key : mlkem.key.Key)
    (pb_ciphertext : Slice U8)
    (pb_agreed_secret : Slice U8)
    (h_key : wfDecapKey pk_mlkem_key params) :
  mlkem.decapsulate 
    pk_mlkem_key
    pb_ciphertext 
    pb_agreed_secret
    ⦃ error pb_agreed_secret' =>
        match error with
        | .NoError =>
            ∃ h_s h_c,
              MLKEM.Decaps params 
                  (pk_mlkem_key.toDecapKey params)
                  (pb_ciphertext.toSpec _ h_c)
                = some (pb_agreed_secret'.toSpec _ h_s)
        | .InvalidArgument =>
            pb_agreed_secret.length ≠ 32 ∨
            pb_ciphertext.length ≠ cipherlength params
        | .MemoryAllocationFailure =>
            out_of_memory
        | _ => False
    ⦄
\end{minted}

The theorem gives a precise post-condition after calling \leaninline{mlkem.decapsulate},
parameterized by the function result \leaninline{error} and the updated value of the output buffer 
\leaninline{pb_agreed_secret'}.
The monadic postcondition notation also asserts that the extracted computation
terminates without panic. Its body separately analyzes the Rust
\rustinline{Error} value returned by the function.

On success, it states functional correctness against FIPS~203
    Algorithm~21, formalized as \leaninline{MLKEM.Decaps}, with matching parameters and arguments.
(This includes implicit rejection:
if re-encryption does not match the ciphertext, the implementation returns the
secret derived as $\overline K = \mathrm{SHAKE256}(z \mathbin\| c)$ rather than
the candidate secret.) 
The specification-level values are explicitly converted from their implementation representations,
which in turn involves auxiliary array length properties \leaninline{h_s} and \leaninline{h_c} 
also provided in this branch of the postcondition. 

The remaining branches precisely characterize length
    errors (the function dynamically checks them on the input and output buffers) and allocation failure
    (the function allocates temporary space, which may fail), and statically exclude every other error code.
For example, the Rust function also checks the key object has its internal \rustinline{has_private_key} flag 
set, which is guaranteed by \leaninline{wfDecapKey} in the theorem.

\bibliographystyle{plain}
\bibliography{paper}
\end{document}